\documentclass[twocolumn,linenumbers]{aastex631}

\shorttitle{M and G dwarfs abundances in the Hyades and Coma Berenices}
\shortauthors{Vilar et al.}

\graphicspath{{./}{figures/}}
\usepackage[utf8]{inputenc}
\usepackage{newunicodechar}

\newunicodechar{−}{\ensuremath{-}}

\begin{document}
\nolinenumbers
\title{Chemical Abundances of M and G Dwarfs in the Hyades and Coma Berenices Open Clusters from APOGEE Spectra}

\correspondingauthor{Deusalete Vilar}
\email{deusavilar@academico.ufs.br}

\author[0009-0007-2704-6198]{Deusalete Vilar}
\affiliation{Departamento de F\'isica, Universidade Federal de Sergipe, Av. Marcelo Deda Chagas, S/N Cep 49.107-230, S\~ao Crist\'ov\~ao, SE, Brazil}

\author[0000-0002-7883-5425]{Diogo Souto}
\affiliation{Departamento de F\'isica, Universidade Federal de Sergipe, Av. Marcelo Deda Chagas, S/N Cep 49.107-230, S\~ao Crist\'ov\~ao, SE, Brazil}

\author[0000-0001-6476-0576]{Katia Cunha}
\affiliation{Observatório Nacional/MCTIC, R. Gen. José Cristino, 77,  20921-400, Rio de Janeiro, Brazil}
\affiliation{Steward Observatory, University of Arizona, 933 North Cherry Avenue, Tucson, AZ 85721-0065, USA}

\author[0009-0000-2228-8234]{Anderson Andrade-Silva}
\affiliation{Departamento de F\'isica, Universidade Federal de Sergipe, Av. Marcelo Deda Chagas, S/N Cep 49.107-230, S\~ao Crist\'ov\~ao, SE, Brazil}

\author[0000-0003-0506-8269]{Veronica Loaiza-Tacuri}
\affiliation{Departamento de F\'isica, Universidade Federal de Sergipe, Av. Marcelo Deda Chagas, S/N Cep 49.107-230, S\~ao Crist\'ov\~ao, SE, Brazil}

\author[0000-0003-0697-2209]{Fábio Wanderley}
\affiliation{Observatório Nacional/MCTIC, R. Gen. José Cristino, 77, 20921-400, Rio de Janeiro, Brazil}

\author[0000-0002-0134-2024]{Verne V. Smith}
\affiliation{NSF’s NOIRLab, 950 N. Cherry Ave. Tucson, AZ 85719 USA}

\author[0000-0002-9612-8054]{Vinicius Grilo}
\affiliation{Departamento de F\'isica, Universidade Federal de Sergipe, Av. Marcelo Deda Chagas, S/N Cep 49.107-230, S\~ao Crist\'ov\~ao, SE, Brazil}

\author[0000-0003-2745-8241]{Cintia F. Martinez}
\affiliation{Instituto de Astronomía y Física del Espacio (CONICET-UBA), C.C. 67 Sucursal 28, C1428EHA, Buenos Aires, Argentina}

\author[0000-0002-0149-1302]{B\'{a}rbara Rojas-Ayala}
\affiliation{Instituto de Alta Investigaci\'on, Universidad de Tarapac\'a, Casilla 7D, Arica, Chile}

\author[0000-0003-0179-9662]{Zachary Way}
\affiliation{Department of Physics and Astronomy, Georgia State University, Atlanta, GA 30302, USA}

\begin{abstract}
\nolinenumbers
Open clusters are one of the best astrophysical laboratories we have available for stellar astrophysics studies. This work presents metallicities and individual abundances for fourteen M dwarfs and six G dwarfs from two well-known open clusters: Hyades and Coma Berenices. Our analysis is based on near-infrared (1.51--1.69 \micron), high-resolution ($R\sim22,500$) spectra obtained from the SDSS IV/APOGEE Survey. Using one-dimensional, plane-parallel MARCS model atmospheres, the APOGEE line list, and the Turbospectrum radiative transfer code in local thermodynamic equilibrium, we derived spectroscopic stellar parameters for the M dwarfs, along with abundances of 13 elements (C, O, Na, Mg, Al, Si, K, Ca, Ti, V, Cr, Mn, and Fe) for both M and G dwarfs.
We find a high degree of chemical homogeneity within each cluster when comparing abundances derived from M and G dwarfs: $\delta$[M/H] (M dwarfs − G dwarfs) of 0.01$\pm$0.04, and 0.02$\pm$0.03 for the Hyades and Coma Berenices, respectively. The overall cluster metallicities derived from M dwarfs (Hyades: 0.16$\pm$0.03 and Coma Berenices: 0.02$\pm$0.06) are consistent with previous literature determinations.
Finally, we demonstrate the value of M dwarfs as key tracers in galactic archaeology, emphasizing their potential for studying galactic metallicity gradients and chemical evolution.
\end{abstract}


\section{Introduction} \label{sec:intro}

Open clusters offer a valuable framework for studying the formation and evolution of our Galaxy. 
Characterizing the individual stellar members within open clusters is crucial for understanding their formation histories, chemical compositions, and dynamical evolution.
Stellar astrophysics studies commonly assume that stars within a given open cluster formed from the same molecular cloud. Consequently, these stars are expected to share similar ages and initial chemical compositions, which are inherited from their natal environment. This intrinsic homogeneity makes open clusters exceptional laboratories for probing stellar properties, particularly chemical abundances (\citealt{desilva2007}, \citealt{bovy2016}, \citealt{Souto2018}, \citealt{poovelil2020}, \citealt{cheng2021}, \citealt{bragaglia2022}, \citealt{sinha2024}). 

Most chemical abundance studies of stars in open clusters have focused on solar-type dwarfs or red giant stars, which, due their higher luminosities, make them more accessible to optical spectrographs. However, stars with effective temperatures ($T_{\rm eff}$) near or above solar values are expected to experience more efficient atomic diffusion processes (\citealt{choi2016}, \citealt{gao2018}, \citealt{Souto2018}, \citealt{Souto2019}), which can alter their surface chemical composition patterns.  Likewise, red giants are subject to internal mixing mechanisms, such as convective dredge-up events (\citealt{nagede}) that alter the original stellar surface abundances, in particular those of C and N, thus modifying the chemical signatures inherited from their natal molecular clouds.
In contrast, low-mass stars, such as M dwarfs, are known to preserve their surface chemical compositions with minimal changes over time, owing to their larger convective envelopes (which inhibit diffusion) and exceptionally long lifespans, which can extend to trillions of years. Despite their numerical dominance in the Galaxy, accounting for approximately 70\% of all stars (\citealt{Salpeter1955}, \citealt{reid1997}, \citealt{kroupa2002}), detailed chemical abundance studies of M dwarfs remain relatively limited compared to those of solar-type or red giant stars.
This scarcity is largely due to observational challenges. M dwarfs have low effective temperatures, resulting in spectra dominated by complex molecular features, such as H$_2$O, TiO, and CO, which make high-precision optical abundance analysis particularly challenging.

The APOGEE (\citealt{Majewski2017}) and CARMENES (\citealt{carmenes2014}) surveys have played a leading role in realizing the chemical and stellar characterization of M-dwarf stars from high-resolution spectroscopy. As part of SDSS-III and IV, the APOGEE survey observed approximately 700,000 stars across the Milky Way with the goal of advancing our understanding of Galactic formation and evolution, with about 3\% of these targets being M dwarfs. The CARMENES survey, in contrast, is dedicated to detecting low-mass exoplanets orbiting M dwarfs and has already obtained spectra for approximately 300 such stars.
Studies based on data from these surveys and other data (\citealt{lindgren2015}, \citealt{caballero2016}, \citealt{Souto2017}, \citealt{baroch2018}, \citealt{passagger2018}, \citealt{reiners2018}, \citealt{cifuentes2020}, \citealt{ishikawa2020}, \citealt{marfil2021},  \citealt{Shan2021}, \citealt{Sarmento2021}, \citealt{Souto2022}, \citealt{Melo2024}, \citealt{Hejazi2024}, and \citealt{Olander2025}) have demonstrated that stellar parameters can be reliably derived from high-resolution M-dwarf spectra, enabling detailed chemical characterization of this most common stellar population in the Galaxy. 

Despite the growing interest in low-mass stars, to date, to our knowledge, there are no high-resolution multi-element abundance studies targeting M dwarf stars in open clusters. 
Using APOGEE spectra, the previous works by \citet{Souto2021} and \citet{Wanderley2023} determined the metallicities, but not the chemical patterns, for M-dwarf members of the Coma Berenices and the Hyades open clusters. 
In particular, Souto et al. (2021) also analyzed the FGK-type dwarfs, finding clear signatures of diffusion for iron abundances in F-dwarf members of Coma Berenices.
Large surveys, such as APOGEE, GALAH \citealt{galah2015} and Gaia ESO \citealt{bragaglia2022}, have also targeted a number of G-type dwarfs in several Galactic open clusters and have studied atomic diffusion and chemical homogeneity in Galactic open clusters \citealt{cunha2016, gao2018, Souto2018, liu2019,  semenova2020, spina2021, grilo2024, sinha2024}.

In this study, we derive the chemical abundance patterns of M-type and G-type dwarf stellar members of two nearby young open clusters: the Hyades and Coma Berenices. These two open clusters are well-established members of the Milky Way thin disk, and their chemical abundances serve as important references for studying chemical evolution of the solar neighborhood and, in particular, the time evolution of the Galactic metallicity gradients. 
By doing a homogeneous spectroscopic analysis of both M-type and G-type dwarfs adopting the same atomic and molecular APOGEE line list (Smith et al. 2021) and radiative transfer code, is especially valuable, as it allows one to evaluate possible systematics in the abundances of the different stellar classes, investigate the degree of chemical homogeneity of the open cluster gas, beyond internal uncertainties, and further testing predictions of diffusion signatures.

The Hyades open cluster, with an estimated age of 625 to 650 Myr (\citealt{perryman1998}; \citealt{lodieu2018}; \citealt{martin2018}; \citealt{arentoft2019}), is also one of the nearest clusters to the Sun at a distance of 47.0 $\pm$ 0.2 pc (\citealt{lodieu2019A}), and with a minimal reddening of $E(B - V) = 0.001$ mag (\citealt{taylor2006}). Several high-resolution optical studies of Hyades members, mostly F, G, K-type stars, have found this young cluster to be slightly metal-rich (\citealt{liu2016}, \citealt{casamiquela2021}, \citealt{wanderley2024}).  
Coma Berenices is another nearby open cluster at a distance of $85 \pm 7.1$ pc (\citealt{tang2018}) and an estimated age between 600 and 800 Myr (\citealt{casewell2006}, \citealt{Kraus2007}, \citealt{casewell2014}, \citealt{tang2018}). 
Due to its proximity to the Sun, the cluster is also subject to minimal interstellar reddening, with $E(B-V) < 0.01$ mag (\citealt{taylor2007}). 

This paper presents, for the first time, detailed individual abundances for M dwarfs belonging to open clusters. We select fourteen M and six G main-sequence dwarf stars belonging to two young open clusters of the Milky Way disk: Hyades and Coma Berenices. We derive stellar parameters and the abundances of thirteen elements: C, O, Na, Mg, Al, Si, K, Ca, Ti, V, Cr, Mn, and Fe, using near-infrared spectra from the APOGEE survey.
This paper is organized as follows: Section \ref{sec:style} presents the APOGEE data and sample selection. Section \ref{sec:modeling} presents the methodology used to determine the chemical abundance of the target stars. Section \ref{sec:cite} discusses the results, and Section \ref{sec:final} summarizes our findings.

\section{Observations and Selected Sample} \label{sec:style}

In this study, we analyze spectra from the SDSS-IV APOGEE Survey \citep{Majewski2017}, using data from the seventeenth public data release (DR17; \citealt{blanton2017}, \citealt{Abdurro'uf2022}). 
The stars analyzed in this study were observed using the APOGEE-N and APOGEE-S cryogenic, multi-fiber spectrographs. Each instrument is equipped with 300 fibers and is mounted on a 2.5-meter-class telescope: the Sloan Foundation Telescope at Apache Point Observatory (APO) in the Northern Hemisphere (\citealt{bowen1973}, \citealt{gunn2006}) and the du Pont Telescope at Las Campanas Observatory in the Southern Hemisphere (\citealt{Wilson2018}), respectively.
The APOGEE spectrographs operate in the H band, covering a spectral range of 1.51 to 1.69 $\micron$ with high-resolution (R $\sim$ 22,500). 
The spectra used in this analysis were processed with the APOGEE Stellar Parameters and Chemical Abundances Pipeline (ASPCAP), an automated reduction pipeline \cite{Nidever2015} and \cite{garcia2016}.

The selection of M dwarfs from the Hyades open cluster was based on the stellar members analyzed for metallicities by \cite{Wanderley2023}. From this sample, we selected seven stars whose spectra have high signal-to-noise ratios (SNR) and well-defined spectral lines. Additionally, four G dwarfs were selected via a cross-match between the APOGEE DR17 database and the catalog of \cite{hunt2023}, requiring membership probabilities greater than 0.5 and radial velocities consistent with the expected cluster value ($39.36 \pm 0.26$ km s$^{-1}$; \citealt{leao2019}). The selected G dwarfs also exhibit well-defined spectral lines and have effective temperatures close to solar values.
For the Coma Berenices open cluster, our sample consists of the same M and G dwarfs previously analyzed by \cite{Souto2021}. 
While \cite{Souto2021} focused on the determination of  metallicities for a sample of F, G, K, and M-type  dwarfs, this study extends that analysis by deriving individual elemental abundances for their M and G dwarfs.

\begin{figure*}
    \centering
    \includegraphics[width=0.49\textwidth]{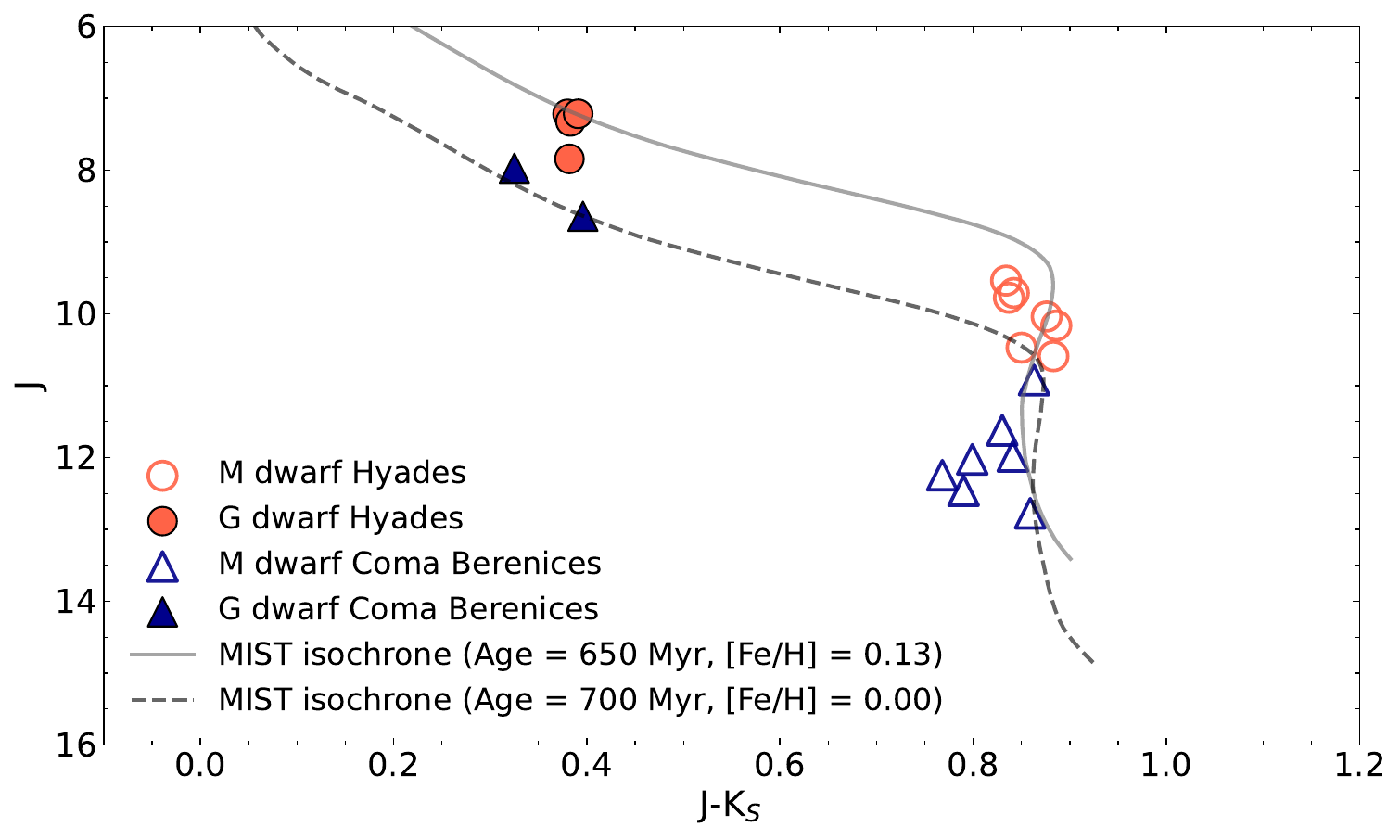}
    \includegraphics[width=0.49\textwidth]{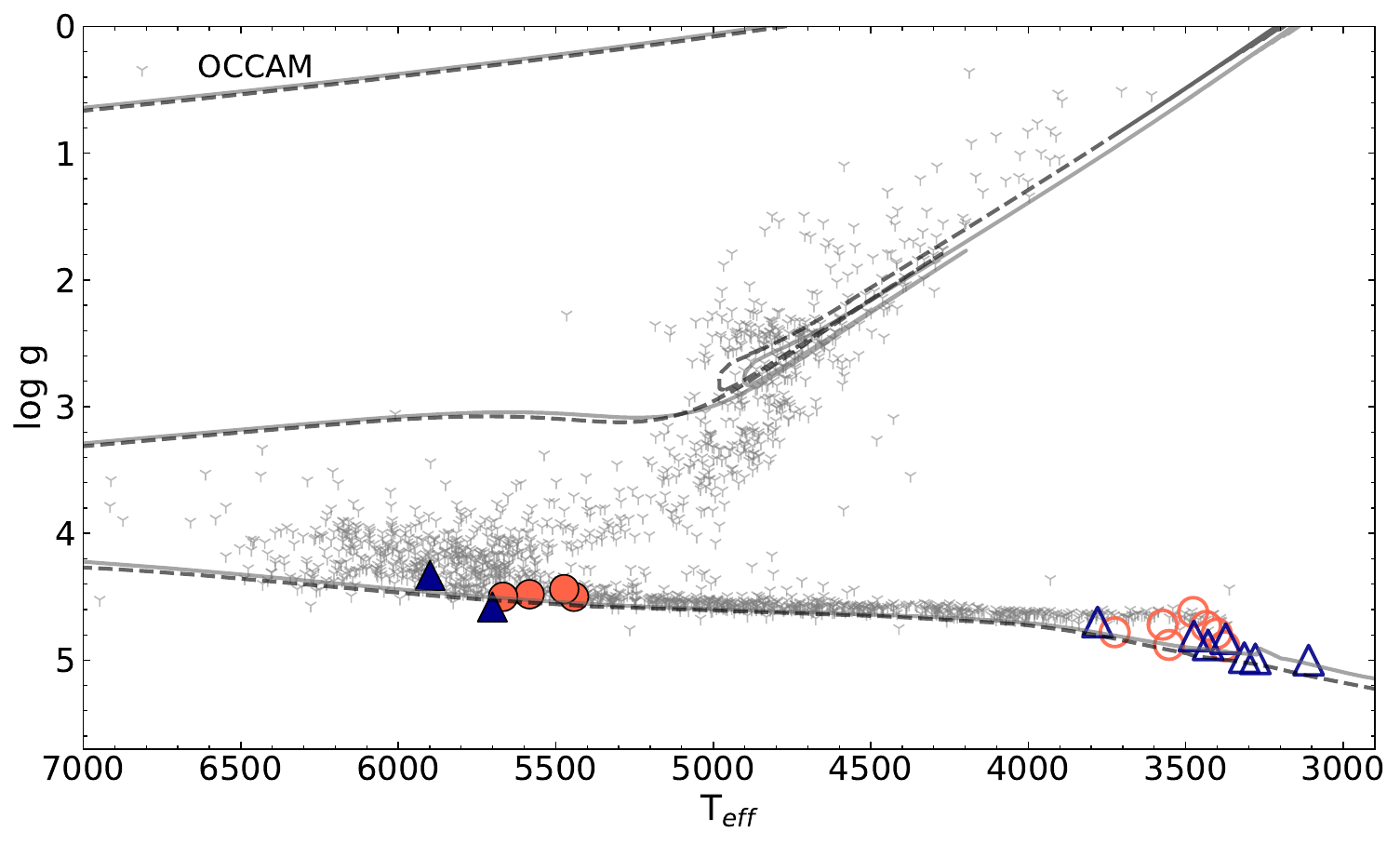}
    \caption{ Left panel: A CMD diagram showing (J–Ks) versus J for our sample of selected stars, with the red circles representing the Hyades, and the blue triangles Coma Berenices. Open symbols represent M and filled symbols represent the G dwarf stars. Two MIST isochrones of different ages are shown: 650 Myr (solid line) and 700 Myr (dashed).
    Right panel: a Kiel diagram for the Hyades and Coma Berenices. Symbols are the same as those in the left panel, but with values from the OCCAM (\citealt{occam}) survey, as gray inverted triangles.}
    \label{fig:cmd}
\end{figure*}

Summarizing, our sample consists of seven M dwarfs and four G dwarfs from the Hyades, as well as seven M dwarfs and two G dwarfs from Coma Berenices. 
Although the sample is relatively small (20 stars), it consists of carefully selected targets that are well-suited for abundance comparisons between M and G dwarfs. The M dwarfs exhibit high SNR, and the G dwarfs have effective temperatures close to solar.
Figure \ref{fig:cmd}, left panel, presents a color-magnitude diagram (CMD) using 2MASS photometry. Open symbols represent M dwarfs, while filled symbols denote G dwarfs. In the diagram, red circles correspond to the Hyades, and blue triangles to Coma Berenices. Superimposed are two MIST isochrones (\citealt{mist}) corresponding to different ages and metallicities representative of each cluster: 650 Myr (\citealt{lodieu2018}; \citealt{martin2018}) with [Fe/H] = 0.13 (solid line) for the Hyades cluster, and 700 Myr (\citealt{casewell2006}, \citealt{Kraus2007}, \citealt{casewell2014}, \citealt{tang2018}) with [Fe/H] = 0.00 (dashed line) for the Coma Berenices cluster. 
Most stars align closely with their respective isochrones, indicating consistency with their expected evolutionary stages. We note that no reddening is adopted as both open clusters are close to the Sun, having $E(B-V) < 0.01$. The complete list of sample stars (and their atmospheric parameters) is provided in Table \ref{tab:sample}.

\begin{deluxetable*}{lrrrrrrr}
\tabletypesize{\scriptsize}
\tablecaption{Stellar parameters}
\tablewidth{0pt}
\label{tab:sample}
\tablehead{
\colhead{APOGEE ID} &
\colhead{J} &
\colhead{H} &
\colhead{K$_{\rm s}$} &
\colhead{$T_{\rm eff}$} &
\colhead{$\log g$} &
\colhead{$v \sin i$} &
\colhead{SNR}
}
\startdata
\hline
\multicolumn{8}{c}{Hyades} \\
\hline
\multicolumn{8}{c}{M dwarfs} \\
2M03413689$+$2320374 & 10.038 & 9.418 & 9.162 & 3435 & 4.73 & 3.14 & 271 \\
2M03415547$+$1845359 & 9.710 & 9.107 & 8.868 & 3573 & 4.72 & 1.79 & 330 \\
2M03591417$+$2202380 & 9.538 & 8.914 & 8.704 & 3725 & 4.78 & 2.36 & 226 \\
2M04172811$+$1454038 & 10.471 & 9.869 & 9.621 & 3401 & 4.79 & 10.60 & 215 \\
2M04363080$+$1905273 & 10.590 & 9.968 & 9.707 & 3378 & 4.89 & 3.92 & 172 \\
2M04363893$+$1836567 & 9.777 & 9.153 & 8.940 & 3553 & 4.88 & 2.38 & 240 \\
2M04544410$+$1940513 & 10.163 & 9.491 & 9.277 & 3476 & 4.62 & 2.77 & 212 \\
\hline
\multicolumn{8}{c}{G dwarfs} \\
2M04473755$+$1815315 & 7.217 & 6.901 & 6.837 & 5583 & 4.48 & 5.14 & 161  \\
2M04293075$+$2640176 & 7.844 & 7.534 & 7.462 & 5442 & 4.50 & 4.38 & 1094 \\
2M04325943$+$1549084 & 7.323 & 7.017 & 6.941 & 5473 & 4.44 & 5.01 & 706  \\
2M04142562$+$1437300 & 7.216 & 6.916 & 6.825 & 5667 & 4.47 & 6.58 & 64 \\
\hline
\multicolumn{8}{c}{Coma Berenices} \\
\hline
\multicolumn{8}{c}{M dwarfs} \\
2M12193796$+$2634445 & 12.776 & 9.374 & 9.264 & 3110 & 5.00 & 42.26 & 130 \\
2M12264027$+$2718434 & 12.462 & 11.896 & 11.672 & 3314 & 4.98 & 5.17 & 192 \\
2M12201448$+$2526072 & 12.246 & 11.705 & 11.477 & 3373 & 4.83 & 14.97 & 138 \\
2M12231356$+$2602185 & 12.025 & 11.519 & 11.226 & 3279 & 4.99 & 2.62 & 513 \\
2M12255421$+$2651387 & 11.984 & 11.389 & 11.143 & 3429 & 4.88 & 3.04 & 387 \\
2M12250262$+$2642382 & 11.621 & 11.028 & 10.791 & 3474 & 4.81 & 3.51 & 295 \\
2M12241121$+$2653166 & 10.921 & 10.265 & 10.058 & 3780 & 4.70 & 9.17 & 199 \\
\hline
\multicolumn{8}{c}{G dwarfs} \\
2M12270627$+$2650445 & 8.642 & 8.327 & 8.246 & 5701 & 4.58 & 5.90 & 1223 \\
2M12204557$+$2545572 & 7.974 & 7.740 & 7.649 & 5899 & 4.33 & 7.00 & 949 \\
\enddata
\end{deluxetable*}

\section{Atmospheric Parameter and Chemical Abundance analyses} \label{sec:modeling}

We performed spectral synthesis modeling of the APOGEE spectra to derive stellar parameters, including effective temperatures, surface gravities ($\log g$), metallicities ([Fe/H]), and individual chemical abundances for the M and G dwarfs in our sample.
The spectral synthesis was conducted using 1D plane-parallel local thermodynamic equilibrium (LTE) MARCS model atmospheres \citep{Gustafsson2008}, the APOGEE DR17 atomic and molecular line list (\citealt{Smith2021}), and the Turbospectrum radiative transfer code (\citealt{AlvarezPlez1998}, \citealt{Plez2012}).
Chemical abundances were derived using the BACCHUS wrapper \citep{Masseron2016} in its semi-automatic mode, which selects the synthetic spectrum that minimizes the chi-squared value relative to the observed spectrum. All spectral fits were visually inspected, and spectral lines exhibiting poor fits were excluded from the final analysis to ensure the robustness and reliability of the derived abundances.

For M dwarfs, individual abundances were determined using 113 spectral lines, when available, following the list provided by \cite{Melo2024}. These lines encompass contributions from various atomic and molecular species, allowing for a comprehensive chemical characterization of M dwarfs.
For G dwarfs, abundances were derived using a total of 71 spectral lines, with 31 of them corresponding to Fe I transitions. The selection of spectral lines for G dwarfs was based on the works of \cite{Souto2018} and \cite{grilo2024}, which provide well-established diagnostic lines appropriate to the spectral analysis of stars within the $T_{\rm eff}$ and metallicity regime of the studied sample.

The $T_{\rm eff}$ and $\log g$ for M dwarfs in the Hyades and Coma Berenices open clusters were determined using oxygen abundances derived from H$_{2}$O and OH spectral lines.
The strengths of the H$_{2}$O and OH features are sensitive to the oxygen abundance; however, the H$_{2}$O lines are highly responsive to changes in $T_{\rm eff}$, while the OH lines are relatively insensitive to these changes. In addition, OH lines are more sensitive to changes in $\log g$, whereas H$_{2}$O lines are less affected. This differential sensitivity enables us to constrain atmospheric parameters by seeking consistency between the abundances derived from the two molecular species. Through this approach, we determined $T_{\rm eff}$, $\log g$, and oxygen abundances for the M dwarfs in our sample, following the methodology described in \cite{Souto2020}.
We note that the Coma Berenices study of \cite{Souto2021} derived $T_{\rm eff}$, $\log g$, and [Fe/H] for using the same methodology as adopted in this work, and those previously derived values were adopted in this study. 

For the G dwarfs, we adopted the spectroscopic $T_{\rm eff}$ values provided by the ASPCAP pipeline (uncalibrated values) from DR17 as a reference.
Surface gravities were calculated using the Stefan-Boltzmann equation (Eq. 1), 

\begin{eqnarray}\begin{array}{rcl}\mathrm{log}g & = & \mathrm{log}{g}_{\odot }+\mathrm{log}\left(\displaystyle \frac{{M}_{\star }}{{M}_{\odot }}\right)+4\mathrm{log}\left(\displaystyle \frac{{T}_{\star }}{{T}_{\odot }}\right)\\ & & +0.4({M}_{{bol},\star }-{M}_{{bol},\odot }).\end{array}\end{eqnarray}

assuming solar reference values ($T_{\rm eff\odot}$ = 5772 K, $\log g_\odot$ = 4.438 dex, and $M_{\rm bol,\odot}$ = 4.75; \citealt{prsa2016}).
Stellar masses, bolometric magnitudes, and luminosities were derived from MIST isochrones \citep{mist, choi2016}, using our spectroscopic $T_{\rm eff}$ and absolute $H$-band magnitudes ($M_{\rm H}$) as reference parameters.
Figure \ref{fig:cmd}, right panel, shows the Kiel diagram ($T_{\rm eff}$ vs. log $g$) for our sample stars. Symbols are the same as in the left panel.

Given the consistent methodology adopted here and in our previous studies, we adopted \cite{Melo2024} as a reference to estimate abundance uncertainties for the M dwarfs and  \cite{Souto2018} for the G dwarfs. Both of these studies assess elemental abundance sensitivities to variations in atmospheric parameters ($T_{\rm eff}$, $\log g$, and microturbulence velocity ($\xi$)) to propagate the errors and estimate abundance uncertainties. In addition, \cite{Melo2024} incorporated the effects of pseudo-continuum placement and signal-to-noise ratio (SNR) uncertainties into their error budget. For our M dwarf sample, the typical abundance uncertainty is 0.11 dex, with Si exhibiting the largest uncertainties, approximately 0.15 dex. As expected, the uncertainties for G dwarfs are generally smaller, reflecting the higher precision in their atmospheric parameter determinations.

\begin{figure*}
    \centering
   { \includegraphics[width=0.79\textwidth]{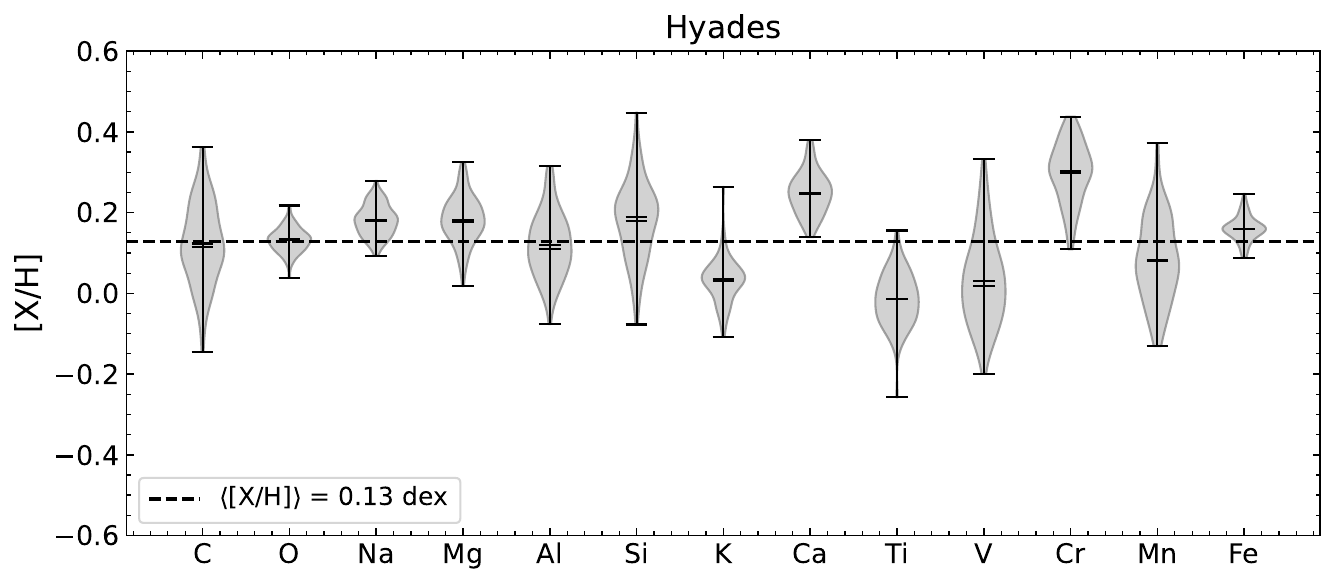}}\\
   { \includegraphics[width=0.79\textwidth]{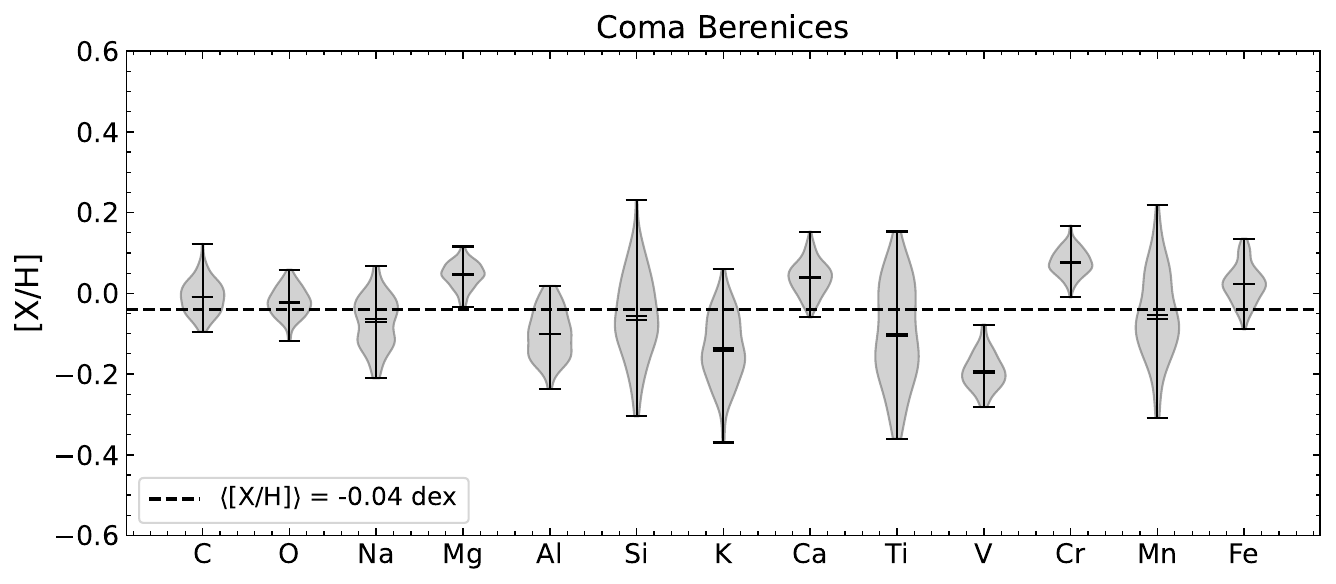}}
    \caption{Elemental abundance distributions for the Hyades (top panel) and Coma Berenices (bottom panel) open clusters, presented as violin diagrams. Each violin shows the distribution of abundances derived for all studied elements, combining both G and M dwarf stars. The black dashed line indicates the overall mean metallicity ([M/H]) considering all studied elements for each cluster.}
    \label{fig:violin_abundances}
\end{figure*}

\section{Results and discussion} \label{sec:cite}

\subsection{Mean Abundances for the Hyades and Coma Berenices}

The individual abundances of elements C, O, Na, Mg, Al, Si, K, Ca, Ti, V, Cr, Mn, and Fe for all studied members of the Hyades and Coma Berenices open clusters are listed in Table \ref{tab:sensitivity}. 
For each star, we provide the mean elemental abundance, the standard deviation, and the abundance uncertainty, separated by $/$. Additionally, at the bottom of each cluster section, we report the mean elemental abundances along with the corresponding standard deviations of the mean, calculated considering all studied stars irrespective of spectral type.

Two values are provided for the Fe abundances of the M dwarfs in the table: one based on Fe I lines and another on FeH lines. In the Hyades, the mean cluster abundance derived using FeH transitions is [Fe/H]=+0.16$\pm$0.03, while the mean from the Fe I lines is $\sim$0.15 dex larger with more scatter (+0.31$\pm$0.07). The systematically larger mean Fe abundance found using Fe I diagnostic lines may be related, in part, to possible effects of magnetic fields on the Fe I lines. \cite{Wanderley2023} have shown that the Hyades M-dwarf stars suffer from radius inflation, which is caused by magnetic fields, and that this correlates with activity indices.  In a subsequent study, \cite{wanderley2024} used Fe I lines in the APOGEE spectra of M-dwarf members of the Pleiades (another young open cluster from the Milky Way disk) to derive mean magnetic fields. Based on the possibility that the Fe I lines are affected in the M-dwarf members of these relatively young open clusters, we will adopt the iron abundances derived from the FeH lines. 
In addition, as will be discussed in the next section, the Fe abundances from FeH provide a better overall match to the G-dwarf members in the Hyades, noting, however, that the systematic difference between the abundances of Fe I and FeH is not as large for the M dwarfs from Coma Berenices: mean [Fe/H]=+0.06$\pm$0.04 from Fe I and [Fe/H]=+0.02$\pm$0.06 from FeH.

In Figure \ref{fig:violin_abundances}, we present the derived abundances for all analyzed elements in the Hyades (top panel) and Coma Berenices (bottom panel) open clusters. The distributions are presented using violin plots, which show both the central tendency and the spread of the data. Each ``violin'' displays the kernel density estimate of the abundance distribution, allowing 
to assess the symmetry, scatter, and typical values for each element. The width of each shape corresponds to the density of abundance values, while internal markers indicate the mean and median. We note that we use the average values obtained from both G and M dwarf stars for each cluster in the figure. The black dashed line indicates the mean overall metallicity for each cluster ($\langle$[M/H]$\rangle$), where we obtain $\langle$[M/H]$\rangle$ = 0.13 $\pm$ 0.03 and -0.04 $\pm$ 0.04 for the Hyades and Coma Berenices, respectively.

\begin{figure*}
    \centering
    \includegraphics[width=0.79\textwidth]{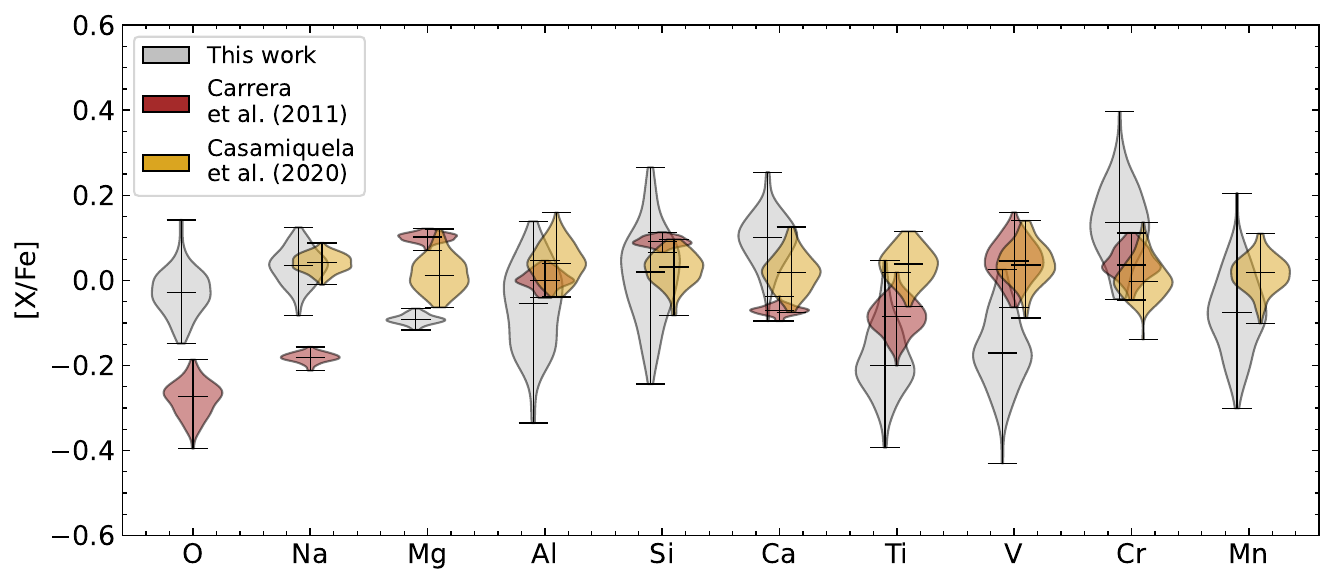}\\     
    \caption{Comparison of chemical abundances between stars in the studied open cluster Hyades (gray) and the works of \cite{carrera2011} (brown) and \cite{casamiquela2020}(yellow).
}
\label{fig:violin_literature}
\end{figure*}

Despite its proximity, the Coma Berenices open cluster lacks extensive high-resolution abundance studies, which limits our ability to perform detailed abundance comparisons with previous studies.
For the Hyades, a comparison between our abundance results and those from the literature is presented in Figure \ref{fig:violin_literature}. The violin plots display the metallicities derived in this work (gray), alongside results from \citet{carrera2011} (brown) and \citet{casamiquela2020} (yellow). Overall, our average metallicities show good agreement with these previous studies, with differences of $\Delta$[M/H] = 0.01 dex relative to \citet{carrera2011} and $\Delta$[M/H] = –0.01 dex relative to \citet{casamiquela2020}. 

Our oxygen abundances were systematically higher, by roughly 0.20 dex, than those estimated for \cite{carrera2011}. One possible explanation for this difference is the challenge in precisely measuring oxygen abundances from the forbidden O I line at 6300 \AA, which suffers from contamination by a Ni I line and by telluric absorption features in \cite{carrera2011} study.
The oxygen abundances of this work are obtained only for the M dwarfs from the OH lines. These lines are temperature sensitive and become weaker as the effective temperature increases for solar-like stars.

Our Mg abundances present a slightly lower value (-0.10 -- -0.20 dex) compared to \citet{casamiquela2020} and \citet{carrera2011}.
Among the iron-peak elements, vanadium displays a larger scatter in our measurements, likely due to the presence of only a single weak V I line in the APOGEE spectral range. Manganese also shows slightly more dispersion, which may be attributed to non-LTE effects that can bias abundance measurements, particularly in Sun-like stars (\citealt{Asplund2021}). 

For the metallicities (assumed as iron abundances), our results for the Hyades are overall consistent within the uncertainties with those from previous studies by \cite{paulson2003}, \cite{dutra2016}, \cite{netopil2016}, \cite{takeda2020}; all these are high-resolution optical analyses.
We find $\langle$[Fe/H]$\rangle$ = 0.16 $\pm$ 0.03 dex for M dwarfs and $\langle$[Fe/H]$\rangle$ = 0.15 $\pm$ 0.02 dex for G dwarfs, as shown in the violin plot (Figure \ref{fig:violin_abundances}).
Our results are also consistent with values from studies that compare stellar properties with theoretical isochrones and evolutionary models (\citealt{choi2016}, \citealt{wolfgang2023}, \citealt{wang2024}, \citealt{gossege2018}), particularly in the M dwarf regime.

\citet{liu2016} studied sixteen solar-type stars (5650 K $<$ $T_{\rm eff}$ $<$ 6250 K) in the Hyades based on high-resolution ($R$=60,000) and high SNR spectra.
Their stellar atmospheric parameters were derived through excitation and ionization equilibrium of Fe I and Fe II lines in a line-by-line differential analysis relative to the Sun, yielding a metallicity of ⟨[Fe/H]⟩ = 0.16 $\pm$ 0.02 dex. These values are consistent with our findings for the Hyades M and G dwarfs. 
\citet{liu2016} interpreted their results as chemical inhomogeneities at the 0.02 dex level. In the present study, we obtain a similarly small standard deviation of the mean, 0.02 dex, for our Fe abundances.

\subsection{Comparing Abundance Results between the M Dwarfs and G Dwarfs}

As discussed previously, having both M-dwarf and G-dwarf open cluster members analyzed homogeneously enables us to assess whether the abundances derived for M dwarfs are consistent with those of G dwarfs, as well as to evaluate the level of chemical homogeneity within each cluster based upon two types of stars having significantly different ranges in $T_{\rm eff}$.
This comparison is particularly relevant, given that the APOGEE line list was calibrated using the solar and Arcturus spectra to refine $\log gf$ values for atomic transitions (\citealt{Smith2021}). 
Overall, our abundance results for the M- and G-type dwarfs reveal a considerable degree of chemical homogeneity across the two studied open clusters. We find a good agreement, within the uncertainties, between M and G dwarf abundances, further validating the methodology employed in this study to derive spectroscopic atmospheric parameters and individual chemical abundances for the studied dwarfs. This assumes that the open cluster abundances are homogeneous to some degree.

\begin{figure*}
\centering
{\includegraphics[width=0.95\textwidth]{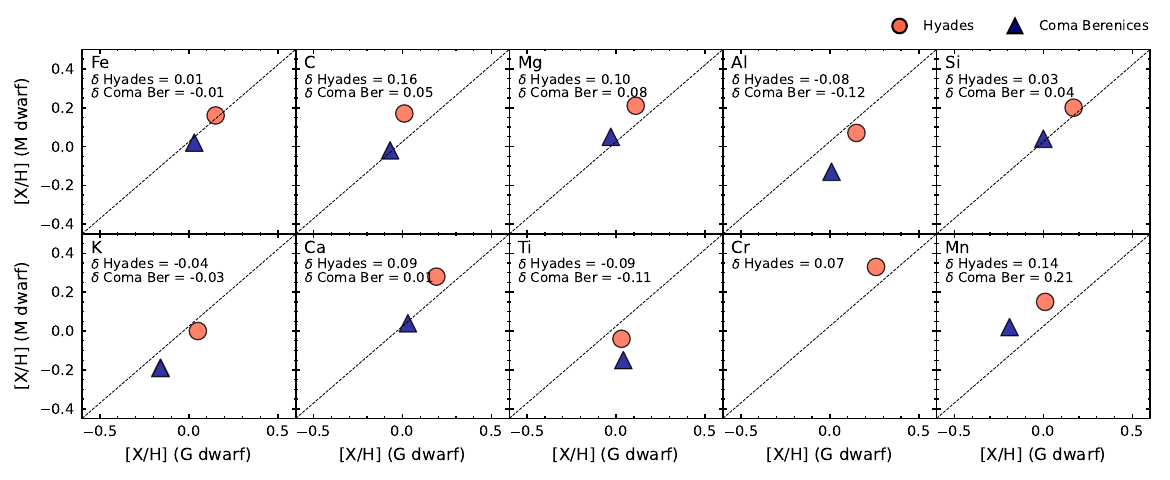}}
\caption{Chemical abundance comparison between M and G dwarf stars in the studied open clusters Hyades and Coma Berenices. The red circles represent the mean abundances for the Hyades and the blue triangles for Coma Berenices. $\delta$ abundance value (G dwarfs – M dwarfs) is indicated in the top left corner of each panel.}
\label{fig:abundance}
\end{figure*}

In Figure \ref{fig:abundance}, we compare the mean chemical abundances derived for the M and G dwarfs for each open cluster. Elemental abundances are shown for Fe, C, Mg, Al, Si, K, Ca, Ti, Cr, and Mn, with each panel in the figure corresponding to one element. The red-filled circles correspond to the mean abundances (and standard deviations) obtained for the Hyades members. In contrast, the filled blue triangles correspond to the mean abundances (and standard deviations) obtained for the Coma Berenices members. 
The mean $\delta$ abundance value (M dwarfs – G dwarfs) is indicated at the top left corner of each panel, and a 1:1 reference line is included to guide the visual comparison. (We note that we excluded Na and V, as we were unable to determine the abundances of these elements for either G or M dwarfs in one of the clusters analyzed.)

Starting with the Hyades, the elements for which there is a small difference (less than or equal to 0.05 dex) between the mean abundances of the M-types and the G-types are: Fe, Mg, K, and Si. (We note that the Si I lines in M dwarfs become badly blended as the effective temperature decreases, resulting in only a few secured abundance measurements.) The mean difference is slightly larger for the elements Al, Ca, Ti, and Cr. The two elements showing a more significant discrepancy in the abundances are C and Mn, for which the differences between the mean abundances of the M dwarfs are higher by 0.16 and 0.14 dex, respectively. 
Finally, the mean abundance difference for all the elements for the Hyades is 0.034 $\pm$ 0.081 dex, a small difference that is well within the typical abundance uncertainties, indicating that the chemical distribution of the analyzed stars is homogeneous to this level.  

A comparison of the individual elemental abundances between M and G dwarfs in Coma Berenices (Figure \ref{fig:abundance}) also reveals good agreement overall, particularly for the elements Fe, C, Si, K, and Ca, all having abundance differences of 0.05 or less. The elements Mg and Cr have slightly higher differences (less than 0.08 dex), while Ti and Mn have differences of -0.11 and -0.21. The mean abundance difference for Coma Berenices in all cases is 0.013 $\pm$0.094 dex, also homogeneous within the uncertainties. 
For the iron abundances in particular, the differences between the abundances of the G and M dwarfs are very small, with $\delta$([Fe/H]$_{\rm Mdwarfs}$ − [Fe/H]$_{\rm Gdwarfs}$) = 0.01, and -0.01 dex for the Hyades and Coma Berenices, respectively.

\begin{figure*}
  \centering
  {\includegraphics[width=0.9\textwidth]{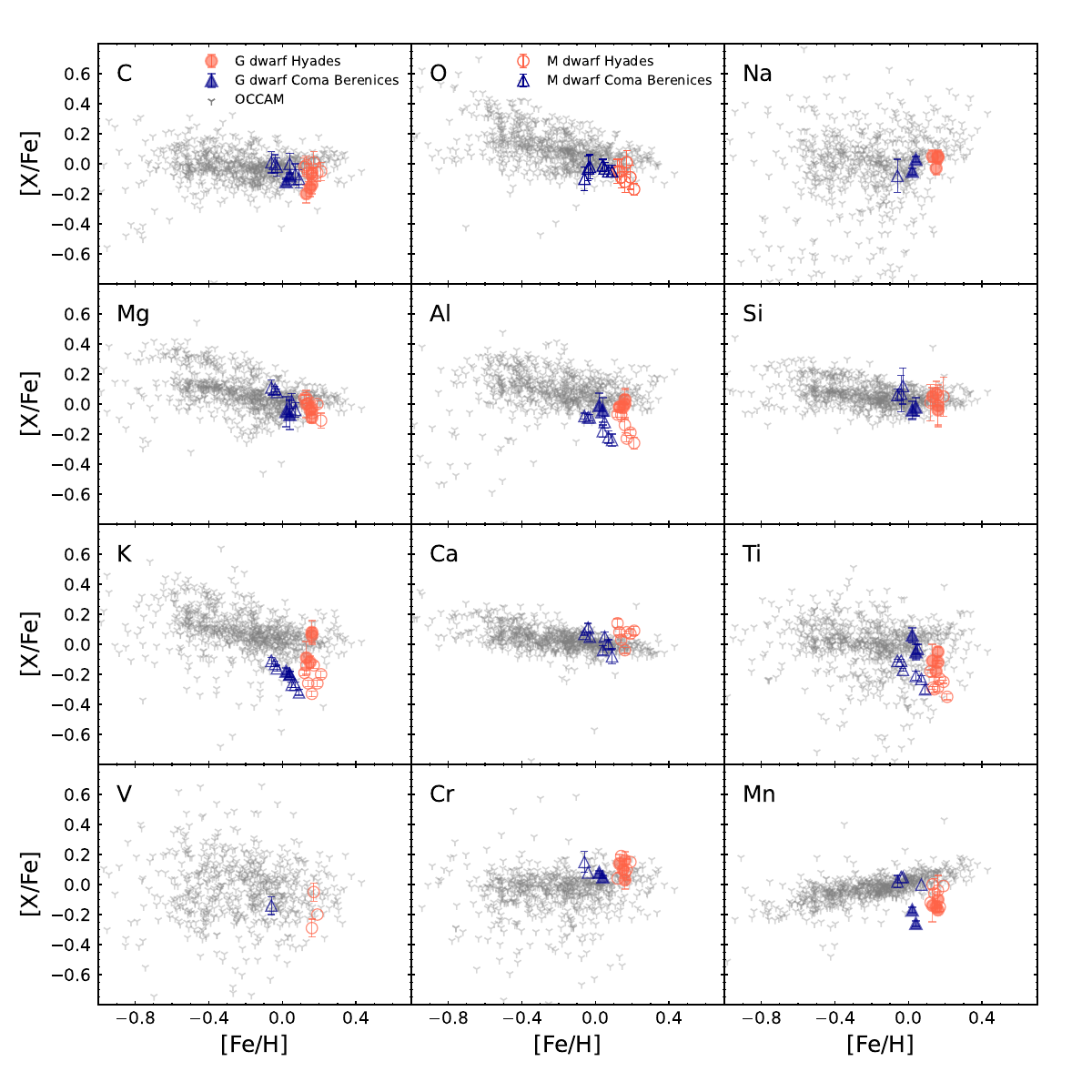}} 
  \caption{Chemical abundance distribution ([X/Fe] vs [Fe/H]) of G and M dwarfs in the open clusters studied in this work, compared to values from the OCCAM DR17 (\citealt{occam}). The red circle represents the Hyades, and the blue triangles are from Coma Berenices, and the gray inverted triangles are results from OCCAM star.}
  \label{fig:metal}
\end{figure*}

\subsection{Results for the Hyades and Coma Berenices in the Context of Other Galactic Open Clusters}

In Figure \ref{fig:metal}, we present the abundance distributions ([X/Fe] vs [Fe/H]) for the G and M dwarf stars in the two open clusters analyzed in this study. The G dwarfs are shown as filled symbols, while the M dwarfs as open symbols, the Hyades and Coma Berenices results are in red and blue respectively.
For comparison, we include results from the Open Cluster Chemical Abundances and Mapping (OCCAM; \citealt{occam}) survey which is also based on APOGEE data, shown as gray inverted triangles.
Overall, our derived abundances follow the distribution of abundances of the open clusters in the OCCAM catalog at above solar metallicities, but for Al, K, Ti, and Mn, these appear to be systematically lower, particularly for M dwarfs (open symbols in the Figure).

Some elements exhibit more significant abundance scatter, such as, Mg, Al, Na, and Mn) as shown in Figure \ref{fig:metal}. These elements display higher standard deviations in their derived abundances, with values of 0.10, 0.13, 0.13, and 0.13 dex, respectively.
Elements such as Mg and Al are particularly sensitive to uncertainties in stellar atmospheric parameters (\citealt{Melo2024}), which naturally leads to larger abundance uncertainties for these species. 
Challenges in deriving accurate Na abundances are related to the presence of only two weak lines within the APOGEE spectral range (\citealt{Souto2022}). 
Additionally, Mn abundances in solar-type stars are known to require non-LTE corrections for accurate determination (\citealt{Asplund2005}, \citealt{Bergemann2019}). The lack of such corrections in our LTE-based analysis may lead to systematic uncertainties in the derived Mn abundances for G dwarfs, contributing to the larger star-to-star abundance scatter and the slight offsets observed when compared to expected trends. 
Titanium exhibits the largest standard deviation of the mean abundance among all elements analyzed in this study (0.13 dex). Previous works have systematically reported lower Ti abundances derived from APOGEE spectra (\citealt{Jonsson2020}, \citealt{Souto2022}), suggesting that the current Ti line list may require further refinement to improve abundance accuracy.

In the Milky Way, abundance gradients provide crucial insights into the chemical evolution of the Galactic disk and can be studied across various stellar populations, including young OB stars, H II regions, Cepheids, planetary nebulae, and red giants, both in the field and within open clusters (\citealt{friel2002}, \citealt{daflon2004}, \citealt{rood2007}, \citealt{magrini2010}, \citealt{carrera2011}, \citealt{frinchaboy2013}, \citealt{genovali2013}, \citealt{magrini2023}).
These gradients reflect the interplay between star formation, nucleosynthesis, and radial migration over cosmic time.
Among these tracers, open clusters serve as compelling tools for investigating the temporal evolution of abundance gradients in the Galactic disk. As chemically homogeneous stellar populations with well-constrained ages and distances, they provide a robust framework for testing chemical evolution models. Recent studies have reinforced their importance in this context, offering new insights into radial metallicity trends and their time evolution (\citealt{frinchaboy2013}, \citealt{cunha2016}, \citealt{casamiquela2021}, \citealt{occam}, \citealt{spina2022}, \citealt{guerco2025}).

The results of this study allow us to investigate abundance gradients using M dwarfs for the first time in the literature.
One of the standouts of M dwarfs is that their low mass allows them to remain nearly chemically unchanged throughout their long lifetimes. Unlike higher-mass stars, where mixing and atomic diffusion can significantly alter surface abundances, M dwarfs experience minimal internal mixing. Consequently, their observed surface abundances provide a more direct record of the chemical composition of their natal molecular clouds, making them ideal tracers of the Galactic metallicity distribution over time from the Galactic archaeology perspective.
Nonetheless, these cool and small stars are intrinsically faint, especially at blue wavelengths, which makes them difficult to observe at large distances with enough quality to determine precise stellar abundances. However, the next generation of extremely large telescopes, equipped with high-resolution near-infrared spectrographs, combined with continued advancements in stellar models and atomic/molecular line lists tailored for cool stars, may enable us to overcome these limitations. This advancement will facilitate the utilization of M dwarfs as indicators of the primordial chemical history of the Galaxy.

\begin{figure}
    \centering
   {\includegraphics[width=0.5\textwidth]{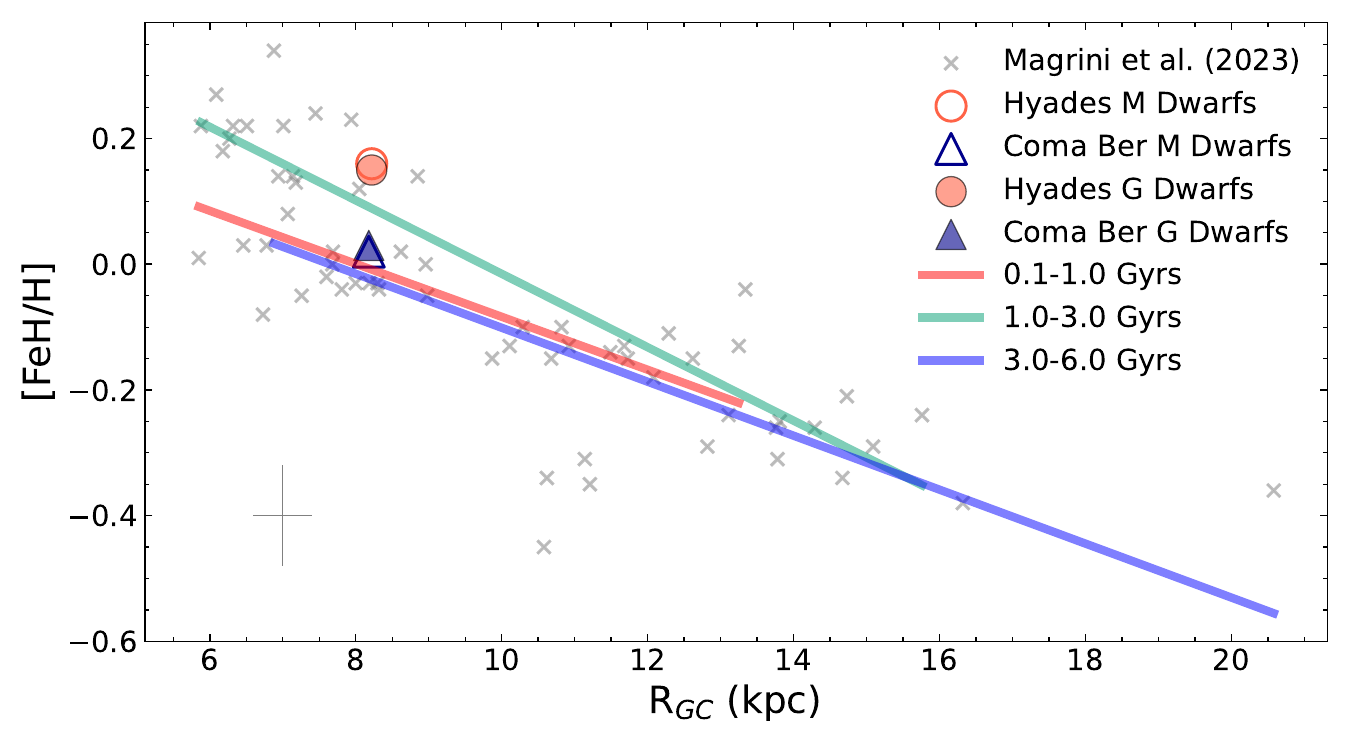}}
\caption{Stellar metallicity as a function of galactocentric distance (kpc). Our metallicity results are represented by solid symbols for G dwarfs and hollow symbols for M dwarfs. The Hyades are red circles, and Coma Berenices are blue triangles, plotted alongside the data from \cite{magrini2023}, shown as small gray Xs. Different lines on the graph represent the abundance gradient slopes for various ages, as described by \cite{magrini2023}.}
\label{fig:kpc}
\end{figure}

Figure \ref{fig:kpc} shows the distribution of metallicity as a function of Galactocentric distance (in kpc) for Galactic open clusters, with data results for stars from different clusters from \cite{magrini2023} represented by gray crosses.
Filled symbols for G dwarfs and open symbols for M dwarfs represent the mean metallicities obtained in this study. The Hyades are red circles, and Coma Berenices are blue triangles, respectively. The symbol for the data for the G dwarfs from Coma Berenices is not visible due to the small difference in metallicity values, only 0.01 dex, between the two spectral classes.
We also include three metallicity gradients based on the discussion in \cite{magrini2023}: the red line represents clusters younger than 1 Gyr, the green line corresponds to clusters with ages between 1 and 3 Gyr, and the blue line denotes clusters older than 3 Gyr. The typical abundance uncertainty is presented in the bottom left panel of the Figure.
Our results for Coma Berenices fall within the metallicity gradients defined by \cite{magrini2023}. The Hyades, however, exhibit a slightly higher metallicity than the average distribution in the solar neighborhood, placing it above the expected metallicity gradients.
We also note that the abundances derived from M and G dwarfs are consistent within the uncertainties.

\section{Conclusions}
\label{sec:final}
This paper presented, for the first time in the literature, individual abundances (C, O, Na, Mg, Al, Si, K, Ca, Ti, V, Cr, Mn, and Fe) for a sample of 14 M dwarf stars spanning a $T_{\rm eff}$ range from $\sim$ 3100 to 3780 K, as well as six G dwarfs, which are $T_{\rm eff}$ about solar, in two open clusters: Hyades and Coma Berenices. 
The abundance analysis is based on 1-D LTE using the SDSS-IV APOGEE infrared spectra.
The overall metallicity ([Fe/H]) for the clusters is: 0.16 $\pm$ 0.03 dex and 0.15 $\pm$ 0.02 dex in the Hyades, and 0.02 $\pm$ 0.06 dex and 0.03 $\pm$ 0.01 dex for the Coma Berenices for M- and G-dwarfs in each cluster, respectively.
Considering the overall abundance content for each open cluster, we obtain [M/H] = 0.14 and -0.04 dex for the M dwarfs of the Hyades and Coma Berenices, and [M/H] = 0.12 and -0.02 dex for the G dwarfs of Hyades and Coma Berenices, respectively.

A comparison with high-resolution abundance studies in the literature for the Hyades yields an excellent agreement, with $\delta$[M/H] $\leq$ $\pm$0.01.
Our comparison between M and G dwarfs in the Hyades reveals a small but consistent metallicity offset of 0.02 dex, with G dwarfs appearing more metal-poor. 
While our results do not statistically confirm the 0.02 dex-level inhomogeneity reported by \citet{liu2016}, the observed offset may support their proposed scenario of pollution of metal-poor gas before complete mixing of the proto-cluster cloud. 
However, another possible explanation is that this difference suggests that warmer stars may experience more efficient atomic diffusion, resulting in lower surface abundances. While our current sample is insufficient to confirm the presence of atomic diffusion in the Hyades conclusively, the observed trend motivates future investigations using a larger sample of stars with different masses in the Hyades.

We also present the first investigation of metallicity gradients in the Milky Way using M dwarf stars. Taking advantage of their chemical stability over long timescales, we demonstrate that M dwarfs serve as reliable tracers of the Galactic metallicity distribution. Our results show consistency with established metallicity gradients derived from other stellar populations. The Coma Berenices clusters follow the trend predicted by recent models, while the Hyades exhibit a slightly higher metallicity than the average distribution in the solar neighborhood.

\section*{acknowledgments}
D.S. acknowledges support from the Foundation for Research and Technological Innovation Support of the State of Sergipe (FAPITEC/SE) and the National Council for Scientific and Technological Development (CNPq), under grant numbers 794017/2013 and 442133/2023-5.
K.C. and V.S. acknowledge support from the National Science Foundation through NSF grant No. AST-2009507.

Funding for the Sloan Digital Sky Survey IV has been provided by the Alfred P. Sloan Foundation, the U.S. Department of Energy Office of Science, and the Participating Institutions. SDSS-IV acknowledges support and resources from the Center for High-Performance Computing at the University of Utah. The SDSS website is www.sdss.org.
SDSS-IV is managed by the Astrophysical Research consortium for the Participating Institutions of the SDSS Collaboration including the Brazilian Participation Group, the Carnegie Institution for Science, Carnegie Mellon University, the Chilean Participation Group, the French Participation Group, Harvard-Smithsonian Center for Astrophysics, Instituto de Astrof\'isica de Canarias, The Johns Hopkins University, Kavli Institute for the Physics and Mathematics of the Universe (IPMU) /  University of Tokyo, Lawrence Berkeley National Laboratory, Leibniz Institut f\"ur Astrophysik Potsdam (AIP),  Max-Planck-Institut f\"ur Astronomie (MPIA Heidelberg), Max-Planck-Institut f\"ur Astrophysik (MPA Garching), Max-Planck-Institut f\"ur Extraterrestrische Physik (MPE), National Astronomical Observatory of China, New Mexico State University, New York University, University of Notre Dame, Observat\'orio Nacional / MCTI, The Ohio State University, Pennsylvania State University, Shanghai Astronomical Observatory, United Kingdom Participation Group, Universidad Nacional Aut\'onoma de M\'exico, University of Arizona, University of Colorado Boulder, University of Oxford, University of Portsmouth, University of Utah, University of Virginia, University of Washington, University of Wisconsin, Vanderbilt University, and Yale University.

{\it Facilities: {Sloan}}.

Software: BACCHUS (\citealt{Masseron2016}), Turbospectrum (\citealt{AlvarezPlez1998}; \citealt{Plez2012}; \href{https://github.com/bertrandplez/Turbospectrum2019}{https://github.com/bertrandplez/Turbospectrum2019}).

\movetabledown=0.7in
\clearpage
\startlongtable
\begin{longrotatetable}
\begin{deluxetable*}{lccccccccccccccc}
\tabletypesize{\tiny}
\tablecaption{Stellar individual abundances}
\tablewidth{0pt}
\label{tab:sensitivity}
\tablehead{
\colhead{APOGEE ID} &
\colhead{[Fe/H]} &
\colhead{[Fe I/H]} &
\colhead{[FeH/H]} &
\colhead{[O/H]} &
\colhead{[C/H]}&
\colhead{[Na/H]} &
\colhead{[Mg/H]} &
\colhead{[Al/H]} &
\colhead{[Si/H]} &
\colhead{[K/H]} &
\colhead{[Ca/H]} &
\colhead{[Ti/H]} &
\colhead{[V/H]} &
\colhead{[Cr/H]} &
\colhead{[Mn/H]} 
}
\startdata
Hyades \\
\hline
M dwarfs\\
2M03413689+2320374 & 0.19/0.02 & 0.32/0.02 & 0.12/0.01 & 0.14/0.08 & 0.17/0.06 & ... & 0.23/0.02 & 0.12/0.02 & 0.20/0.12 & 0.00/0.02 & 0.33/0.03 & 0.01/0.01 & ... & 0.23 & 0.07 \\
2M03415547+1845359 & 0.24/0.03 & 0.36/0.06 & 0.16/0.02 & 0.12/0.07 & 0.16/0.06 & ... & 0.18/0.01 & 0.10/0.05 & 0.30/0.05 & -0.09/0.02 & 0.20/0.02 & -0.05/0.02 & -0.05/0.06 & 0.32 & 0.20/0.1\\
2M03591417+2202380 & 0.16/0.03 & 0.16/0.02 & 0.16/0.04 & 0.17/0.08 & 0.17/0.07 & ... & -0.14/0.01 & -0.07/0.02 & -0.02/0.04 & 0.02/0.02 & 0.24/0.01 & -0.07/0.01 & 0.11/0.06 & 0.26/0.07 & 0.01/0.03\\
2M04172811+1454038 & 0.17/0.02 & 0.30/0.03 & 0.13/0.03 & 0.12/0.08 & 0.16/0.06 & ... & 0.21/0.02 & 0.15/0.02 & ... & 0.03/0.02 & 0.25/0.02 & -0.12/0.01 & ... & ... & 0.17\\
2M04363080+1905273 & 0.24/0.03 & 0.32/0.05 & 0.21/0.02 & 0.07/0.04 & 0.19/0.06 & ... & 0.13/0.05 & -0.02/0.04 & ... & 0.04/0.02 & 0.33/0.03 & -0.11/0.02 & ... & ... & ... \\
2M04363893+1836567 & 0.24/0.04 & 0.38/0.06 & 0.14/0.02 & 0.15/0.07 & 0.18/0.06 & ... & 0.24/0.01 & 0.17/0.05 & ... & -0.02/0.02 & 0.27/0.02 & -0.06/0.02 & -0.13/0.08 & 0.43 & 0.18/0.1\\
2M04544410+1940513 & 0.25/0.02 & 0.31/0.03 & 0.19/0.02 & 0.16/0.08 & 0.19/0.06 & ... & 0.25/0.02 & 0.06/0.01 & 0.30/0.12 & -0.01/0.02 & 0.32/0.03 & 0.00/0.01 & 0.05 & 0.40 & 0.24\\
Mean  & 0.21 & 0.31 & 0.16 & 0.13 & 0.17 & ... & 0.23 & 0.11 & 0.20 & 0.00 & 0.28 & -0.04 & -0.02 & 0.32 & 0.18\\
Std & 0.04 & 0.07 & 0.03 & 0.03 & 0.01 & ... & 0.18 & 0.11 & 0.15 & 0.04 & 0.05 & 0.07 & 0.09 & 0.08 & 0.10\\
G dwarfs\\
2M04473755+1815315 & 0.16/0.05 & ... & 0.02/0.05 & 0.21 & 0.13/0.02 & 0.19/0.02 & 0.13/0.1 & 0.08/0.08 & 0.23/0.06 & 0.05/0.02 & ... & 0.19/0.05 & -0.01 \\
2M04293075+2640176 & 0.15/0.05 & ... & -0.01/0.05 & 0.12 & 0.11/0.02 & 0.14/0.02 & 0.22/0.1 & 0.04/0.08 & 0.19/0.06 & -0.03/0.02 & ... & 0.25/0.05 & 0.00 \\
2M04325943+1549084 & 0.13/0.05 & ... & -0.07/0.05 & 0.18 & 0.13/0.02 & 0.11/0.02 & 0.17/0.1 & 0.04/0.08 & 0.18/0.06 & 0.01/0.02 & ... & 0.27/0.05 & -0.01 \\
2M04142562+1437300 & 0.16/0.05 & ... & 0.10/0.05 & 0.20 & 0.07/0.02 & 0.17/0.02 & 0.15/0.1 & 0.19/0.08 & 0.31/0.06 & 0.28/0.02 & 0.80/0.04 & 0.34/0.05 & 0.11\\
Mean  & 0.15 & ... & -0.02 & 0.13 & 0.16 & 0.19 & 0.18 & 0.12 & 0.25 & 0.17 & 0.60 & 0.31 & 0.06\\
Std & 0.02 & ... & 0.08 & 0.14 & 0.04 & 0.07 & 0.04 & 0.05 & 0.04 & 0.08 & 0.24 & 0.03 & 0.04\\
\hline
Coma Berenices \\
\hline
M dwarfs\\
2M12193796+2634445 & 0.03/0.04 & ... & 0.05/0.02 & 0.00/0.04 & ... & ... & 0.07/0.05 & -0.07/0.04 & ... & -0.22/0.02 & 0.10/0.03 & 0.02/0.03 & ... & ... & ...\\
2M12264027+2718434 & 0.07/0.03 & 0.07/0.05 & 0.07/0.02 & 0.02/0.04 & 0.00/0.07 & ... & 0.03/0.05 & -0.15/0.04 & ... & -0.20/0.02 & 0.07/0.03 & -0.16/0.02 & ... & ... & 0.07\\
2M12201448+2526072 & 0.03/0.04 & 0.02/0.06 & 0.04/0.02 & 0.02/0.04 & 0.04/0.07 & ... & ... & -0.14/0.04 & ... & -0.16/0.02 & 0.00/0.03 & -0.17/0.03 & ... & ... & ...\\
2M12231356+2602185 & 0.08/0.03 & 0.07/0.04 & 0.09/0.02 & 0.03/0.04 & -0.01/0.06 & ... & 0.07/0.04 & -0.15/0.04 & ... & -0.23/0.02 & 0.01/0.05 & -0.21/0.03 & ... & ... & ... \\
2M12255421+2651387 & -0.01/0.02 & 0.02/0.02 & -0.03/0.02 & -0.02/0.08 & -0.05/0.06 & ... & ... & -0.12/0.02 & ... & -0.19/0.02 & 0.02/0.03 & 0.20/0.01 & ... & ... & 0.02\\
2M12250262+2642382 & 0.01/0.02 & 0.06/0.02 & -0.04/0.02 & -0.02/0.08 & -0.04/0.06 & ... & 0.06/0.02 & -0.13/0.02 & 0.03/0.1 & -0.17/0.02 & 0.07/0.03 & -0.15 & ... & ... & 0.01\\
2M12241121+2653166 & 0.04/0.03 & 0.14/0.02 & -0.06/0.04 & -0.06/0.08 & 0.01/0.07 & -0.14/0.1 & 0.05/0.05 & -0.14/0.02 & 0.00/0.04 & -0.17/0.02 & 0.01/0.01 & -0.17/0.01 & ... & ... & -0.04/0.04\\
Mean &  0.04 & 0.06 & 0.02 & -0.01 & -0.02 & ... & 0.05 & -0.13 & 0.04 & -0.19 & 0.04 & -0.15 & ... & ... & 0.02 \\
Std &   & 0.03 & 0.04 & 0.06 & ... & 0.02 & 0.03 & 0.05 & 0.03 & 0.04 & 0.08 & ... & ... & 0.05 \\
G dwarfs\\
2M12270627+2650445 & 0.02/0.05 & ... & -0.09/0.05 & -0.03 & -0.03/0.02 & 0.01/0.02 & -0.02/0.1 & -0.16/0.08 & 0.05/0.06 & 0.08/0.02 & ... & 0.10/0.05 & -0.15\\
2M12204557+2545572 & 0.04/0.05 & ... & -0.05/0.05 & 0.07 & -0.03/0.02 & 0.00/0.02 & 0.02/0.1 & -0.16/0.08 & 0.01/0.06 & -0.01/0.02 & ... & 0.09/0.05 & -0.22\\
Mean & 0.03 & ... & -0.07 & 0.02 & -0.03 & 0.01 & 0.00 & -0.16 & 0.03 & 0.04 & ... & 0.10 & -0.19 \\
Std &  0.01 & ... & 0.04 & 0.07 & 0.00 & 0.01 & 0.03 & 0.00 & 0.03 & 0.06 & ... & 0.01 & 0.05 \\
\hline
\tablewidth{0pt}	
\enddata
\end{deluxetable*}
\end{longrotatetable}

\vspace{5mm}

\appendix

\label{lastpage}

{}


\begin{thebibliography}{}
\bibitem[Abdurro'uf et al.(2022)]{Abdurro'uf2022} Abdurro'uf, Accetta, K., Aerts, C., et al.\ 2022, \apjs, 259, 2, 35. doi:10.3847/1538-4365/ac4414

\bibitem[Alvarez \& Plez(1998)]{AlvarezPlez1998} Alvarez, R. \& Plez, B.\ 1998, \aap, 330, 1109. doi:10.48550/arXiv.astro-ph/9710157

\bibitem[Arentoft et al.(2019)]{arentoft2019} Arentoft, T., Grundahl, F., White, T.~R., et al.\ 2019, \aap, 622, A190. doi:10.1051/0004-6361/201834690

\bibitem[Asplund(2005)]{Asplund2005} Asplund, M.\ 2005, \araa, 43, 1, 481. doi:10.1146/annurev.astro.42.053102.134001

\bibitem[Asplund et al.(2021)]{Asplund2021} Asplund, M., Amarsi, A.~M., \& Grevesse, N.\ 2021, \aap, 653, A141. doi:10.1051/0004-6361/202140445

\bibitem[Baroch et al.(2018)]{baroch2018} Baroch, D., Morales, J.~C., Ribas, I., et al.\ 2018, \aap, 619, A32. doi:10.1051/0004-6361/201833440

\bibitem[Bergemann et al.(2019)]{Bergemann2019} Bergemann, M., Gallagher, A.~J., Eitner, P., et al.\ 2019, \aap, 631, A80. doi:10.1051/0004-6361/201935811

\bibitem[Blanton et al.(2017)]{blanton2017} Blanton, M.~R., Bershady, M.~A., Abolfathi, B., et al.\ 2017, \aj, 154, 1, 28. doi:10.3847/1538-3881/aa7567

\bibitem[Brandner et al.(2023)]{wolfgang2023} Brandner, W., Calissendorff, P., \& Kopytova, T.\ 2023, \mnras, 518, 1, 662. doi:10.1093/mnras/stac2247

\bibitem[Bragaglia et al.(2022)]{bragaglia2022} Bragaglia, A., Alfaro, E.~J., Flaccomio, E., et al.\ 2022, \aap, 659, A200. doi:10.1051/0004-6361/202142674

\bibitem[Bovy(2016)]{bovy2016} Bovy, J.\ 2016, \apj, 817, 1, 49. doi:10.3847/0004-637X/817/1/49

\bibitem[Bowen \& Vaughan(1973)]{bowen1973} Bowen, I.~S. \& Vaughan, A.~H.\ 1973, \ao, 12, 1430. doi:10.1364/AO.12.001430

\bibitem[Caballero et al.(2016)]{caballero2016} Caballero, J.~A., Cort{\'e}s-Contreras, M., Alonso-Floriano, F.~J., et al.\ 2016, 19th Cambridge Workshop on Cool Stars, Stellar Systems, and the Sun (CS19), 148. doi:10.5281/zenodo.60060

\bibitem[Carrera \& Pancino(2011)]{carrera2011} Carrera, R. \& Pancino, E.\ 2011, \aap, 535, A30. doi:10.1051/0004-6361/201117473

\bibitem[Casamiquela et al.(2020)]{casamiquela2020} Casamiquela, L., Tarricq, Y., Soubiran, C., et al.\ 2020, \aap, 635, A8. doi:10.1051/0004-6361/201936978

\bibitem[Casamiquela et al.(2021)]{casamiquela2021} Casamiquela, L., Castro-Ginard, A., Anders, F., et al.\ 2021, \aap, 654, A151. doi:10.1051/0004-6361/202141779

\bibitem[Casewell et al.(2006)]{casewell2006} Casewell, S.~L., Jameson, R.~F., \& Dobbie, P.~D.\ 2006, \mnras, 365, 2, 447. doi:10.1111/j.1365-2966.2005.09689.x

\bibitem[Casewell et al.(2014)]{casewell2014} Casewell, S.~L., Littlefair, S.~P., Burleigh, M.~R., et al.\ 2014, \mnras, 441, 3, 2644. doi:10.1093/mnras/stu746

\bibitem[Cheng et al.(2021)]{cheng2021} Cheng, C.~M., Price-Jones, N., \& Bovy, J.\ 2021, \mnras, 506, 4, 5573. doi:10.1093/mnras/stab2106

\bibitem[Cifuentes et al.(2020)]{cifuentes2020} Cifuentes, C., Caballero, J.~A., Cort{\'e}s-Contreras, M., et al.\ 2020, \aap, 642, A115. doi:10.1051/0004-6361/202038295

\bibitem[Choi et al.(2016)]{choi2016} Choi, J., Dotter, A., Conroy, C., et al.\ 2016, \apj, 823, 2, 102. doi:10.3847/0004-637X/823/2/102

\bibitem[Cunha et al.(2016)]{cunha2016} Cunha, K., Frinchaboy, P.~M., Souto, D., et al.\ 2016, Astronomische Nachrichten, 337, 8-9, 922. doi:10.1002/asna.201612398

\bibitem[Daflon \& Cunha(2004)]{daflon2004} Daflon, S., \& Cunha, K.\ 2004, \apj, 617, 1115. doi:10.1086/425607

\bibitem[De Silva et al.(2007)]{desilva2007} De Silva, G.~M., Freeman, K.~C., Asplund, M., et al.\ 2007, \aj, 133, 1161. doi:10.1086/511182

\bibitem[De Silva et al.(2015)]{galah2015} De Silva, G.~M., Freeman, K.~C., Bland-Hawthorn, J., et al.\ 2015, \mnras, 449, 2604. doi:10.1093/mnras/stv327

\bibitem[Dotter(2016)]{mist} Dotter, A.\ 2016, \apjs, 222, 8. doi:10.3847/0067-0049/222/1/8

\bibitem[Dutra-Ferreira et al.(2016)]{dutra2016} Dutra-Ferreira, L., Pasquini, L., Smiljanic, R., et al.\ 2016, \aap, 585, A75. doi:10.1051/0004-6361/201526783

\bibitem[Friel et al.(2002)]{friel2002} Friel, E.~D., Janes, K.~A., Tavarez, M., et al.\ 2002, \aj, 124, 2693. doi:10.1086/344161

\bibitem[Frinchaboy et al.(2013)]{frinchaboy2013} Frinchaboy, P.~M., Thompson, B., Jackson, K.~M., et al.\ 2013, \apjl, 777, L1. doi:10.1088/2041-8205/777/1/L1

\bibitem[Garc{\'\i}a P{\'e}rez et al.(2016)]{garcia2016} Garc{\'\i}a P{\'e}rez, A.~E., Allende Prieto, C., Holtzman, J.~A., et al.\ 2016, \aj, 151, 144. doi:10.3847/0004-6256/151/6/144

\bibitem[Gao et al.(2018)]{gao2018} Gao, X., Lind, K., Amarsi, A.~M., et al.\ 2018, \mnras, 481, 2666. doi:10.1093/mnras/sty2414

\bibitem[Genovali et al.(2013)]{genovali2013} Genovali, K., Lemasle, B., Bono, G., et al. 2013, \aap, 554, A132,  doi:10.1051/0004-6361/201321650

\bibitem[Gossage et al.(2018)]{gossege2018} Gossage, S., Conroy, C., Dotter, A., et al. 2018, \apj, 863, 67, doi:10.3847/1538-4357/aad0a0

\bibitem[Guerço et al.(2025)]{guerco2025} Guerço, R., Souto, D., Fernández-Trincado, J. G., et al. 2025, arXiv e-prints, arXiv:2506.19936, doi:10.48550/arXiv.2506.19936

\bibitem[Grilo et al.(2024)]{grilo2024} Grilo, V., Souto, D., Cunha, K., et al. 2024, \mnras, 534, 3005, doi:10.1093/mnras/stae2209

\bibitem[Gunn et al.(2006)]{gunn2006} Gunn, J. E., Siegmund, W. A., Mannery, E. J., et al. 2006, \aj, 131, 2332, doi:10.1086/500975

\bibitem[Gustafsson et al.(2008)]{Gustafsson2008} Gustafsson, B., Edvardsson, B., Eriksson, K., et al. 2008, \aap, 486, 951, doi:10.1051/0004-6361:200809724

\bibitem[Hejazi et al.(2024)]{Hejazi2024} Hejazi, N., Crossfield, I.~J.~M., Souto, D., et al.\ 2024, \apj, 973, 1, 31. doi:10.3847/1538-4357/ad61dc

\bibitem[Hunt \& Reffert(2023)]{hunt2023} Hunt, E. L., \& Reffert, S. 2023, \aap, 673, A114, doi:10.1051/0004-6361/202346285

\bibitem[Ishikawa et al.(2020)]{ishikawa2020} Ishikawa, H. T., Aoki, W., Kotani, T., et al. 2020, \pasj, 72, 102, doi:10.1093/pasj/psaa101

\bibitem[Ishikawa et al.(2022)]{ishikawa2022} Ishikawa, H. T., Aoki, W., Hirano, T., et al. 2022, \aj, 163, 72, doi:10.3847/1538-3881/ac3ee0

\bibitem[Jönsson et al.(2020)]{Jonsson2020} Jönsson, H., Holtzman, J. A., Allende Prieto, C., et al. 2020, \aj, 160, 120, doi:10.3847/1538-3881/aba592

\bibitem[Kraus \& Hillenbrand(2007)]{Kraus2007} Kraus, A. L., \& Hillenbrand, L. A. 2007, \aj, 134, 2340, doi:10.1086/522831

\bibitem[Kroupa(2002)]{kroupa2002} Kroupa, P. 2002, Science, 295, 82, doi:10.1126/science.1067524

\bibitem[Leão et al.(2019)]{leao2019} Leão, I. C., Pasquini, L., Ludwig, H.-G., \& de Medeiros, J. R. 2019, \mnras, 483, 5026, doi:10.1093/mnras/sty3215

\bibitem[Lindgren \& Heiter(2015)]{lindgren2015} Lindgren, S., \& Heiter, U. 2015, AAS/Division for Extreme Solar Systems Abstracts, 47, 115.04

\bibitem[Liu et al.(2016)]{liu2016} Liu, F., Yong, D., Asplund, M., et al. 2016, \mnras, 457, 3934, doi:10.1093/mnras/stw247

\bibitem[Liu et al.(2019)]{liu2019} Liu, F., Asplund, M., Yong, D., et al. 2019, \aap, 627, A117, doi:10.1051/0004-6361/201935306

\bibitem[Lodieu et al.(2018)]{lodieu2018} Lodieu, N., Rebolo, R., \& Pérez-Garrido, A. 2018, \aap, 615, L12, doi:10.1051/0004-6361/201832748

\bibitem[Lodieu et al.(2019)]{lodieu2019A} Lodieu, N., Pérez-Garrido, A., Smart, R. L., \& Silvotti, R. 2019, \aap, 628, A66, doi:10.1051/0004-6361/201935533

\bibitem[Magrini et al.(2010)]{magrini2010} Magrini, L., Randich, S., Zoccali, M., et al. 2010, \aap, 523, A11, doi:10.1051/0004-6361/201015395

\bibitem[Magrini et al.(2023)]{magrini2023} Magrini, L., Viscasillas Vázquez, C., Spina, L., et al. 2023, \aap, 669, A119, doi:10.1051/0004-6361/202244957

\bibitem[Majewski et al.(2017)]{Majewski2017} Majewski, S. R., Schiavon, R. P., Frinchaboy, P. M., et al. 2017, \aj, 154, 94, doi:10.3847/1538-3881/aa784d

\bibitem[Marfil et al.(2021)]{marfil2021} Marfil, E., Tabernero, H. M., Montes, D., et al. 2021, \aap, 656, A162, doi:10.1051/0004-6361/202141980

\bibitem[Mart{\'\i}n et al.(2018)]{martin2018} Mart{\'\i}n, E.~L., Lodieu, N., Pavlenko, Y., \& B{\'e}jar, V.~J.~S.\ 2018, \apj, 856, 40, doi:10.3847/1538-357/aaaeb8

\bibitem[Masseron et al.(2016)]{Masseron2016} Masseron, T., Merle, T., \& Hawkins, K.\ 2016, Astrophysics Source Code Library, ascl:1605.004

\bibitem[Melo et al.(2024)]{Melo2024} Melo, E., Souto, D., Cunha, K., Smith, V.~V., Wanderley, F., Grilo, V., Camara, D., et al.\ 2024, \apj, 973, 90, doi:10.3847/1538-4357/ad5004

\bibitem[Myers et al.(2022)]{occam} Myers, N., Donor, J., Spoo, T., Frinchaboy, P.~M., Cunha, K., Price-Whelan, A.~M., Majewski, S.~R., et al.\ 2022, \aj, 164, 85, doi:10.3847/1538-3881/ac7ce5

\bibitem[Lagarde et al.(2015)]{nagede} Lagarde, N., Miglio, A., Eggenberger, P., Morel, T., Montalb{\'a}n, J., Mosser, B., Rodrigues, T.~S., et al.\ 2015, \aap, 580, A141, doi:10.1051/0004-6361/201525856

\bibitem[Netopil et al.(2016)]{netopil2016} Netopil, M., Paunzen, E., Heiter, U., \& Soubiran, C.\ 2016, \aap, 585, A150, doi:10.1051/0004-6361/201526370

\bibitem[Nidever et al.(2015)]{Nidever2015} Nidever, D.~L., Holtzman, J.~A., Allende Prieto, C., Beland, S., Bender, C., Bizyaev, D., Burton, A., et al.\ 2015, \aj, 150, 173, doi:10.1088/0004-6256/150/6/173

\bibitem[Passegger et al.(2018)]{passagger2018} Passegger, V.~M., Reiners, A., Jeffers, S.~V., Wende-von Berg, S., Sch{\"o}fer, P., Caballero, J.~A., Schweitzer, A., et al.\ 2018, \aap, 615, A6, doi:10.1051/0004-6361/201732312

\bibitem[Paulson et al.(2003)]{paulson2003} Paulson, D.~B., Sneden, C., \& Cochran, W.~D.\ 2003, \aj, 125, 3185, doi:10.1086/375209

\bibitem[Paxton et al.(2011)]{mesa} Paxton, B., Bildsten, L., Dotter, A., Herwig, F., Lesaffre, P., \& Timmes, F.\ 2011, \apjs, 192, 3, doi:10.1088/0067-0049/192/1/3

\bibitem[Perryman et al.(1998)]{perryman1998} Perryman, M.~A.~C., Brown, A.~G.~A., Lebreton, Y., Gomez, A., Turon, C., Cayrel de Strobel, G., Mermilliod, J.~C., et al.\ 1998, \aap, 331, 81, doi:10.48550/arXiv.astro-ph/9707253

\bibitem[Plez(2012)]{Plez2012} Plez, B.\ 2012, Astrophysics Source Code Library, ascl:1205.004

\bibitem[Pr{\v{s}}a et al.(2016)]{prsa2016} Pr{\v{s}}a, A., Harmanec, P., Torres, G., Mamajek, E., Asplund, M., Capitaine, N., Christensen-Dalsgaard, J., et al.\ 2016, \aj, 152, 41, doi:10.3847/0004-6256/152/2/41

\bibitem[Poovelil et al.(2020)]{poovelil2020} Poovelil, V.~J., Zasowski, G., Hasselquist, S., et al.\ 2020, \apj, 903, 1, 55. doi:10.3847/1538-4357/abb93e

\bibitem[Quirrenbach et al.(2014)]{carmenes2014} Quirrenbach, A., Amado, P.~J., Caballero, J.~A., et al.\ 2014, \procspie, 9147, 91471F. doi:10.1117/12.2056453

\bibitem[Reid \& Gizis(1997)]{reid1997} Reid, I.~N. \& Gizis, J.~E.\ 1997, \aj, 114, 1992. doi:10.1086/118620

\bibitem[Reiners et al.(2018)]{reiners2018} Reiners, A., Zechmeister, M., Caballero, J.~A., et al.\ 2018, \aap, 612, A49. doi:10.1051/0004-6361/201732054

\bibitem[Rood et al.(2007)]{rood2007} Rood, R.~T., Quireza, C., Bania, T.~M., et al.\ 2007, From Stars to Galaxies: Building the Pieces to Build Up the Universe, 374, 169. 

\bibitem[Salpeter(1955)]{Salpeter1955} Salpeter, E.~E.\ 1955, \apj, 121, 161. doi:10.1086/145971

\bibitem[Sarmento et al.(2021)]{Sarmento2021} Sarmento, P., Rojas-Ayala, B., Delgado Mena, E., et al.\ 2021, \aap, 649, A147. doi:10.1051/0004-6361/202039703

\bibitem[Semenova et al.(2020)]{semenova2020} Semenova, E., Bergemann, M., Deal, M., et al.\ 2020, \aap, 643, A164. doi:10.1051/0004-6361/202038833

\bibitem[Shan et al.(2021)]{Shan2021} Shan, Y., Reiners, A., Fabbian, D., et al.\ 2021, \aap, 654, A118. doi:10.1051/0004-6361/202141530

\bibitem[Sinha et al.(2024)]{sinha2024} Sinha, A., Zasowski, G., Frinchaboy, P., et al.\ 2024, \apj, 975, 1, 89. doi:10.3847/1538-4357/ad78e1

\bibitem[Smith et al.(2021)]{Smith2021} Smith, V.~V., Bizyaev, D., Cunha, K., et al.\ 2021, \aj, 161, 6, 254. doi:10.3847/1538-3881/abefdc

\bibitem[Souto et al.(2016)]{Souto2016} Souto, D., Cunha, K., Smith, V., et al.\ 2016, \apj, 830, 1, 35. doi:10.3847/0004-637X/830/1/35

\bibitem[Souto et al.(2017)]{Souto2017} Souto, D., Cunha, K., Garc{\'\i}a-Hern{\'a}ndez, D.~A., et al.\ 2017, \apj, 835, 2, 239. doi:10.3847/1538-4357/835/2/239

\bibitem[Souto et al.(2018)]{Souto2018} Souto, D., Cunha, K., Smith, V.~V., et al.\ 2018, \apj, 857, 1, 14. doi:10.3847/1538-4357/aab612

\bibitem[Souto et al.(2019)]{Souto2019} Souto, D., Allende Prieto, C., Cunha, K., et al.\ 2019, \apj, 874, 1, 97. doi:10.3847/1538-4357/ab0b43

\bibitem[Souto et al.(2020)]{Souto2020} Souto, D., Cunha, K., Smith, V.~V., et al.\ 2020, \apj, 890, 2, 133. doi:10.3847/1538-4357/ab6d07

\bibitem[Souto et al.(2021)]{Souto2021} Souto, D., Cunha, K., \& Smith, V.~V.\ 2021, \apj, 917, 1, 11. doi:10.3847/1538-4357/abfdb5

\bibitem[Souto et al.(2022)]{Souto2022} Souto, D., Cunha, K., Smith, V.~V., et al.\ 2022, \apj, 927, 1, 123. doi:10.3847/1538-4357/ac4891

\bibitem[Spina et al.(2018)]{spina2018} Spina, L., Mel{\'e}ndez, J., Casey, A.~R., et al.\ 2018, \apj, 863, 2, 179. doi:10.3847/1538-4357/aad190

\bibitem[Spina et al.(2021)]{spina2021} Spina, L., Ting, Y.-S., De Silva, G.~M., et al.\ 2021, \mnras, 503, 3, 3279. doi:10.1093/mnras/stab471

\bibitem[Spina et al.(2022)]{spina2022} Spina, L., Magrini, L., \& Cunha, K.\ 2022, Universe, 8, 2, 87. doi:10.3390/universe8020087

\bibitem[Takeda \& Honda(2020)]{takeda2020} Takeda, Y. \& Honda, S.\ 2020, \aj, 159, 4, 174. doi:10.3847/1538-3881/ab799f

\bibitem[Tang et al.(2018)]{tang2018} Tang, S.-Y., Chen, W.~P., Chiang, P.~S., et al.\ 2018, \apj, 862, 2, 106. doi:10.3847/1538-4357/aacb7a

\bibitem[Taylor(2006)]{taylor2006} Taylor, B.~J.\ 2006, \aj, 132, 6, 2453. doi:10.1086/508610

\bibitem[Taylor et al.(2008)]{taylor2007} Taylor, B.~J., Joner, M.~D., \& Jeffery, E.~J.\ 2008, \apjs, 176, 1, 262. doi:10.1086/526427

\bibitem[Taylor(2008)]{taylor2008} Taylor, B.~J.\ 2008, \aj, 136, 3, 1388. doi:10.1088/0004-6256/136/3/1388

\bibitem[Wanderley et al.(2023)]{Wanderley2023} Wanderley, F., Cunha, K., Souto, D., et al.\ 2023, \apj, 951, 2, 90. doi:10.3847/1538-4357/acd4bd

\bibitem[Wanderley et al.(2024)]{wanderley2024} Wanderley, F., Cunha, K., Kochukhov, O., et al.\ 2024, \apj, 971, 1, 112. doi:10.3847/1538-4357/ad571f

\bibitem[Wang et al.(2025)]{wang2024} Wang, F., Fang, M., Fu, X., et al.\ 2025, \apj, 979, 1, 92. doi:10.3847/1538-4357/ad960a

\bibitem[Wilson et al.(2018)]{Wilson2018} Wilson, R.~F., Teske, J., Majewski, S.~R., et al.\ 2018, \aj, 155, 2, 68. doi:10.3847/1538-3881/aa9f27

\bibitem[Olander et al.(2025)]{Olander2025} Olander, T., Heiter, U., Piskunov, N., et al.\ 2025, \aap, 698, A289. doi:10.1051/0004-6361/202453476



\end{thebibliography}
\end{document}